\documentclass[12pt,a4paper]{article}

\usepackage{latexsym}
\usepackage{amsfonts}
\usepackage{amssymb}
\usepackage{amsmath}
\usepackage{theorem}
\usepackage{enumerate}
\usepackage{cite}
\usepackage{url}
\usepackage[dvips]{color}
\usepackage{graphicx}

\theoremstyle{plain}
\theorembodyfont{\itshape}
\newtheorem{thm}{Theorem}

\theorembodyfont{\upshape}

\newcommand{\proof}{\noindent {\bf Proof:} \hspace{0.1in}}
\newcommand{\qed}{\hfill\mbox{\raggedright $\Box$}\medskip}
\newcommand{\mydate}{
 \ifcase\month \or
 January\or February\or March\or April\or May\or June\or
 July\or August\or September\or October\or November\or December\fi
 \space \number\year}

\begin{document}

\title{General Solutions to Static Plane Symmetric Einstein's Equations}
\author{L. G. Gomes}
\date{Instituto de Matem\'atica e Computa\c{c}\~ao \\
      Universidade Federal de Itajub\'a , Brazil}
\maketitle

\thispagestyle{empty}

\begin{abstract}
\noindent
A general formula for the metric as an explicit function of the generic energy-momentum tensor 
is given which satisfies static plane symmetric Einstein's equations with cosmological 
constant $\Lambda$. 
In order to illustrate it, the solutions for the vacuum with cosmological 
constant, the perfect fluid with a linear equation of state and the electrically charged plane are 
derived and compared with known results. The general solution with a 
linear relation among the energy-momentum tensor components is also obtained.
\end{abstract}

\section*{Introduction}

 A static plane symmetric spacetime belongs to a class of Lorentzian manifolds
that, to our purposes here, is locally characterized by the existence of a  
coordinate system $(t,x,y,z)$ in which the metric is represented as 
\begin{equation}\label{eq:PlaneMetric}
 g = g_{00}(z) \, dt^2 + g_{11}(z) \, (dx^2 + dy^2) + g_{33}(z) \, dz^2 \quad.
\end{equation}
The non-metric fields must be as well symmetric (\cite{SKMHH}\,). Particularly, in this coordinate system
the energy-momentum tensor is represented as
\begin{equation}\label{eq:GeneralEMT}
(\, T^{\mu}_{\nu} \, ) = \text{\bf diag} \, \{\, \rho (z), -p(z), -p(z), -q(z) \, \} \quad .   
\end{equation}
The Einstein's equations with a cosmological constant $\Lambda$, %
\footnote{In this representation, $\frac{8 \pi \, G}{c^4} = 1$ and the energy density 
is given in units of $[Lenght]^{-2}$}
\begin{equation}\label{eq:GeneralSolutionEFE}
R^{\mu}_{\nu} -  \frac{1}{2} R \, \delta^{\mu}_{\nu} - \Lambda \, \delta^{\mu}_{\nu} =
 T^{\mu}_{\nu}  \quad ,
\end{equation}
together with the equations coming from the system to be considered, establish the dynamics
which connect the six unknown functions: $g_{00}$, $g_{11}$, $g_{33}$, $\rho$, $p$ and $q$. 

A common procedure in dealing with such equations is to choose a coordinate function
$u$ for which $du = g_{33}(z) dz$, and hence, reducing our problem to finding five unknown
functions. If, in addition, we define %
\footnote{This is a further simplification for the form of the metric 
appearing in \cite{DolgKhri}. }
\begin{equation}
 \varphi(u) = \frac{3}{4} \, \frac{d}{du}\, \ln \,|\, g_{11}\,|  
 \qquad \text{and} \qquad
 \psi(u) =  \frac{d}{du}\, \left( \, \frac{1}{2} \, \ln \, g_{00} 
         +  \frac{1}{4} \, \ln \,|\, g_{11}\,| \, \right)  \quad ,
\end{equation}
then the metric (\ref{eq:PlaneMetric}) is written as
\begin{equation}\label{eq:SolutionMetricDogov}
 g = e^{{2 \over 3} \int du \,( 3 \psi(u) - \varphi(u))} \, dt^2
 - e^{{4 \over 3} \int \,du \, \varphi(u)} \, \left(dx^2 + dy^2 \right) - \, du^2 \quad . 
\end{equation}
Since each diagonal component of $T^\mu_\nu$ behaves as a function under any transformation
like $z=z(u)$, Einstein's equations are reduced to 
\begin{eqnarray}\label{Eq:EinsteinDolgov1}
G^t_t = T^t_t +  \Lambda & : & 
- \frac{4}{3} \left( \varphi' + \varphi^2 \right) = \rho +  \Lambda \quad \\
\label{Eq:EinsteinDolgov2}
G^t_t - 4 G^x_x = T^t_t - 4 T^x_x - 3 \Lambda & : & 
4 \left( \psi' + \psi^2 \right) = \rho + 4 p - 3 \Lambda \quad \\
\label{Eq:EinsteinDolgov3}
G^u_u = T^u_u +  \Lambda & : & 
- \frac{4}{3} \, \varphi \, \psi = - q  +  \Lambda \quad .
\end{eqnarray}
These are simple enough to lead us to a general formula for the metric tensor 
as explicit functions of the components $\rho$, $p$ and $q$ of the 
energy-momentum tensor, as we show in the next sections.   

 We proceed as follows: in the first section we give the generic solution of Einstein's
field equations as stated in theorem \ref{thm:GeneralSolution}. Exemplifying its statement,
a general solution is found for any energy-momentum tensor satisfying linear relations among their
components (see eq. \ref{eq:RelacaoLinear}). As special cases, the general vacuum 
($\Lambda \neq 0$), perfect fluid and Einstein-Maxwell solutions are given and compared 
with known results. In section 2 we find the general solutions for the non-generic
energy-momentum tensor. In theorem \ref{thm:GeneralSolution2} we deal with the case $q=\Lambda$ 
while in theorem \ref{thm:GeneralSolution3} with $\rho = - p = q - 2 \Lambda$. In the last
section we conclude that any solution of static plane symmetric Einstein's equations has a local
representation just as stated in at least one of these three theorems.


\section{The Generic Solution}

 The general energy-momentum tensor (\ref{eq:GeneralEMT}) is said to be {\bf generic} if 
there is a space-time point with coordinate $z=z_0$ such that
\begin{equation}\label{eq:GeneralSolutionCond1}
 \big[\,\, q_0 \neq \Lambda \,\,\big] 
 \qquad \,\text{and}\, \qquad
 \big[\,\, \rho_0 \neq q_0 - 2 \Lambda  \quad \text{or} \quad 
 \rho_0 + 4 p_0 + 3 q_0 \neq 6 \Lambda \,\, \big] \quad ,
\end{equation}
where by $\rho_0, p_0 , q_0$ we take the functions $\rho(z), p(z) , q(z)$ evaluated at $z_0$.  
We assume, after a suitable translation, that $z_0=0$. When such conditions are satisfied, 
it is possible to choose constants $z_S$ and $b$ for which 
\begin{equation}\label{eq:zS}
 \frac{q_0 - \Lambda}{z_S} < 0 \qquad , \qquad b>0 
\end{equation}
and 
\begin{equation}\label{eq:b}
 3\, z_S \, (\, \rho_0 - q_0 + 2 \Lambda \, ) \neq b \, (\, \rho_0 + 4 p_0 + 3 q_0 -6 \Lambda \, ) \quad .
\end{equation}
The constant $z_S$ represents the location for a (at least coordinate) singularity in spacetime, 
while arbitrariness of $b$ stands for a choice of the $z$-coordinate scale. In what follows
we will set $b=1$.

\begin{thm}[The Generic Solution ]~ \label{thm:GeneralSolution} \\
For a generic $T_{\mu}^{\nu}=\text{\bf diag} \, \{\, \rho (z), -p(z), -p(z), -q(z) \, \}$, 
there is a maximal open interval 
$I_0$ containing $z=0$ where the function%
\footnote{Indeed, $\Phi = \Phi[\rho, p, q](z)$.}
\begin{equation}\label{eq:GeneralSolutionPhi}
  \Phi (z) = \frac{q - \Lambda}{z -z_S} \, \left( 
 \frac{1}{ 3 (z-z_S) (\rho- q + 2 \Lambda) +  \rho + 4 p + 3 q -6 \Lambda } \right)
\end{equation}
is well defined and 
\begin{equation}\label{eq:GeneralSolutionq-Lambda}
\frac{q(z) - \Lambda}{ z - z_S} > 0  \qquad . 
\end{equation}
%
Furthermore, the metric  
\begin{equation}\label{eq:GeneralSolutionMetric}
 g = e^{2 \int_0^z dz' \, \left( 3 (z'-z_S) - 1 \right) \Phi(z')} \, dt^2
 - e^{4 \int_0^z  dz' \, \Phi(z')} \, \left(dx^2 + dy^2 \right)
 -  \frac{ 12 \, (z-z_S) }{q(z) - \Lambda} \, \Phi(z)^2 \, dz^2 \quad ,
\end{equation}
defined for $z \in I_0$, satisfies the Einstein's equations with cosmological constant
$\Lambda$ ,
\begin{equation}\label{eq:GeneralSolutionEFE2}
R^{\mu}_{\nu} -  \frac{1}{2} R \, \delta^{\mu}_{\nu} - \Lambda \, \delta^{\mu}_{\nu} =
 T^{\mu}_{\nu}  \quad ,
\end{equation}
provided the energy-momentum tensor is conserved
\begin{equation}\label{eq:GeneralSolutionConsEMT}
 \nabla_\mu \, T^{\mu}_\nu  = 0 \quad \iff \quad
 \frac{dq}{dz}= \left[ \, (1-3(z-z_S))\,(\rho + q)  + 4 (p-q) \,\right] \, \Phi \quad .
\end{equation}
\end{thm}
%
%
\proof
To prove this theorem we proceed in a straightforward way and compute the Einstein tensor
of the metric (\ref{eq:GeneralSolutionMetric}). Its Levi-Civita connection has as the only
non-vanishing independent components
\begin{equation}
 \Gamma^t_{tz}= (3(z-z_S)-1)\Phi   \quad ; \quad  \Gamma^x_{xz}= \Gamma^y_{yz}= 2 \Phi
\end{equation}
\begin{equation}
  \Gamma^z_{tt} = \frac{1}{12}\, (3(z-z_S)-1)\, \frac{q-\Lambda}{(z-z_S) \, \Phi} 
\, e^{2 \int dz \, \left( 3 (z-z_S) - 1 \right) \Phi} 
\end{equation}
\begin{equation}
\Gamma^z_{xx} = \Gamma^z_{yy} = - \frac{1}{6}\, \frac{q-\Lambda}{(z-z_S) \, \Phi} 
\, e^{4 \int dz \, \Phi} 
\quad ; \quad
\Gamma^z_{zz} =  \frac{1}{2}\, \left(\frac{1}{(z-z_S)} + \frac{2}{\Phi} \, \frac{d\Phi}{dz} -
\frac{1}{q - \Lambda}  \, \frac{dq}{dz} \, \right)
\end{equation}
The energy-momentum tensor conservation equations, $\nabla_\mu \, T^{\mu}_\nu =0$, 
vanish identically except for $\nu = 3$, the case which is given by equation
(\ref{eq:GeneralSolutionConsEMT}), as one can easily verify.  Using this relation,
the independent components of the curvature tensor 
\begin{equation}
R^{\lambda \mu}_{\kappa \nu} = g^{\mu \mu'}\, R^{\lambda}_{\mu' \kappa \nu} = 
g^{\mu \mu'}\, ( \partial_\kappa \, \Gamma_{\nu \mu'}^\lambda 
-  \partial_\nu \, \Gamma_{\kappa \mu'}^\lambda 
+ \Gamma_{\kappa \sigma}^\lambda \, \Gamma_{\nu \mu'}^\sigma  
- \Gamma_{\nu \sigma}^\lambda \, \Gamma_{\kappa \mu'}^\sigma  )
\end{equation}
are
\begin{equation}
 R^{tx}_{tx} = R^{ty}_{ty} = \frac{3(z-z_S)-1}{6 (z-z_S)} \left( q-\Lambda \right) 
 \quad ; \quad
 R^{tz}_{tz} = \frac{1}{2} \left(\rho - q +2p \right) + \frac{q-\Lambda}{3 (z-z_S)} 
\end{equation}
\begin{equation}
 R^{xy}_{xy} = \frac{q-\Lambda}{3 (z-z_S)}  
 \quad ; \quad
 R^{xz}_{xz} = R^{yz}_{yz} = - \frac{1}{2} \left(\rho + \Lambda \right) - 
 \frac{q-\Lambda}{6 (z-z_S)} 
\end{equation}
The Ricci tensor $R^{\mu}_{\nu} = R^{\lambda \mu}_{\lambda \nu}$ turns out to be a diagonal matrix 
with entries
\begin{eqnarray*}
R^t_t =  \frac{1}{2} \left(\rho + q + 2p - 2 \Lambda \right) \quad 
R^x_x = R^y_y =  - \frac{1}{2} \left(\rho - q + 2 \Lambda \right) \quad
R^z_z =  - \frac{1}{2} \left(\rho + q - 2p + 2 \Lambda \right) \, . 
\end{eqnarray*}
Finally, the curvature scalar is
\begin{equation}
 R  = - T^\mu_\mu - 4 \Lambda = - (\rho - q - 2p) - 4 \Lambda
\end{equation}
Hence Einstein's equations hold for $g$, as can be readily verified.\\[3mm]
\qed
%
%
\vspace{2mm}

To illustrate theorem \ref{thm:GeneralSolution}, let us consider a system for which the components of 
the energy-momentum tensor are related by
\begin{equation}\label{eq:RelacaoLinear}
 \rho + \Lambda = \beta_0 \, (q - \Lambda)
 \qquad \text{and} \qquad 
 p - \Lambda = \beta_1 \, (q - \Lambda)  \quad ,
\end{equation}
with $\beta_0$ and $\beta_1$ constants.
The generic condition upon $T_\nu^\mu$ is expressed here as   
\begin{equation}\label{eq:LinearCondition}
 [\, q_0 \neq \Lambda \, ]  \quad \text{and} \quad
 [\, \beta_0 \neq 1  \quad \text{or} \quad \beta_1 \neq -1 \, ]   \quad .
\end{equation}
The solution with $q = \Lambda$ will be easily obtained with the help of theorem 
\ref{thm:GeneralSolution2}, while the case $\beta_0 = - \beta_1 = 1 $ is integrated in 
theorem \ref{thm:GeneralSolution3}, both to be presented in the next section. 
Therefore we assume throughout this section the generic condition holds and apply 
theorem~\ref{thm:GeneralSolution}. 

The function $\Phi$ defined in (\ref{eq:GeneralSolutionPhi}) is  written as
\begin{equation}\label{eq:phiAffine1}
  \Phi = \frac{1}{(z -a) \,  r(z)} 
\end{equation}
where 
\begin{equation}\label{eq:rGamma}
 r(z)  =  3 (z - z_S)\, (\beta_0 -1 ) + \gamma  \qquad \text{and} \qquad
 \gamma  =  \beta_0 + 4 \beta_1 + 3  \quad .
\end{equation}
The energy-momentum conservation equation (\ref{eq:GeneralSolutionConsEMT}) is readily 
integrated to give
\begin{equation}\label{eq:LinearSolutionConsEMT}
 \ln \, \left( \frac {q(z) - \Lambda}{q_0 - \Lambda} \right)
 = \, - \, \int_0^z \, dz' \,\, 
    \frac{3 (z' - z_S)\, (\beta_0 + 1) - \gamma + 6}{ (z'-z_S) \, r(z')}
\end{equation}
We are left with the three following possibilities:
\begin{enumerate}
\item[{\bf (i)}]{\bf \, Linear System I  ($\gamma \neq 0$ , $\beta_0 \neq 1$) :}
For this condition we obtain
\begin{equation}\label{eq:qAffine1}
 q = \Lambda + \, (q_0 - \Lambda) 
 \, \left( \frac{r(z)}{r(0)} \right)^{- 2 - \frac{2}{\beta_0 -1} + \frac{6}{\gamma}}
 \, \left( 1 - \frac{z}{z_S} \right)^{1 - \frac{6}{\gamma}} 
\end{equation}
 and the metric $g$ is given by formula (\ref{eq:GeneralSolutionMetric}) as
\begin{equation}\label{eq:MetricAffine1}
 \begin{array}{c}
 g = \left( \frac{r(z)}{r(0)} \right)^{ \frac{2}{\beta_0 -1} + \frac{2}{\gamma}}
 \, \left( 1 - \frac{z}{z_S} \right)^{-\frac{2}{\gamma}} \,  \, dt^2 
 -
 \, \left( \frac{r(z)}{r(0)} \right)^{-\frac{4}{\gamma}}
 \, \left( 1 - \frac{z}{z_S} \right)^{\frac{4}{\gamma}}  \, (dx^2 + dy^2) \\[6mm]
 -
 \, \frac{12}{z_s \, (\Lambda - q_0)\, r(0)^2} 
 \, \left( \frac{r(z)}{r(0)} \right)^{\frac{2}{\beta_0 -1} - \frac{6}{\gamma}}
 \, \left( 1 - \frac{z}{z_S} \right)^{-2 + \frac{6}{\gamma}}  \, dz^2
 \end{array} 
\end{equation}
where $z_S$ is chosen such that $r(0)= -3 z_S \, (\beta_0 -1 ) + \gamma \neq 0$ 
and $z_S \,  (\Lambda - q_0) > 0$ .

The family of {\bf vacuum solutions with cosmological constant} $\Lambda \neq 0$ 
(\cite{NoHors}, \cite{Hors}, \cite{SKMHH}) is 
obtained if we set $\rho = p = q = 0$. In order to keep the relations
(\ref{eq:RelacaoLinear}) consistent, we also set $\beta_0=-1$ and $\gamma = 6$.
Here $q_0 = 0$ and $z_S \, \Lambda >0$. Specializing the metric (\ref{eq:MetricAffine1})
in this context, we have 
\begin{eqnarray}\label{eq:MetricAffine1Vacuum1}
 g_{Vacuum} &=& \left( 1 - \frac{z}{z_S+1} \right)^{ - \frac{2}{3}}
 \, \left( 1 - \frac{z}{z_S} \right)^{-\frac{1}{3}} \,  \, dt^2 \nonumber \\
 &-&
 \, \left(1 - \frac{z}{z_S+1} \right)^{-\frac{2}{3}}
 \, \left( 1 - \frac{z}{z_S} \right)^{\frac{2}{3}}  \, (dx^2 + dy^2) \\
 &-&
 \, \frac{1}{3 \, z_s \, \Lambda \, (1+z_S)^2} 
 \, \left( 1 - \frac{z}{z_S+1} \right)^{- 2}
 \, \left( 1 - \frac{z}{z_S} \right)^{-1}  \, dz^2 \quad .\nonumber
\end{eqnarray} 
If $\Lambda > 0$, we can choose $z_S = 1$ 
and define the coordinate $w$ through 
\begin{equation}
  z = 1 - \tan^2(a w)   \qquad \text{,} \quad a = \frac{\sqrt{3 \Lambda}}{2} \quad .
\end{equation}
Except for rescaling the coordinates $t, x, y$ with suitable constant parameters, the metrics 
(\ref{eq:MetricAffine1Vacuum1}) in the $w$-coordinate representation and  
the vacuum solution  presented by Novitn\'y and Horsk\'y  (\cite{NoHors}) are equal to (\cite{SKMHH})
\begin{equation}\label{eq:MetricAffine1Vacuum2}
 g_{Vacuum} =   \cos^2(a w) \, \sin^{-\, \frac{2}{3}}(a w)  \, dt^2 
 -
 \,   \sin^{\frac{4}{3}}(a w)  \, (dx^2 + dy^2) 
 - \, dw^2 \quad .
\end{equation}
 A similar formula is obtained for $\Lambda < 0$ if we change the trigonometric 
functions to  hyperbolic ones and choose  $z_S<0$.

The solution of the {\bf Einstein-Maxwell} problem for a charged infinite plane 
(~\cite{McV}, \cite{AmGr}~) is given if we
take $\Lambda = 0$ and $\rho = p = - q$ . In this case 
\begin{equation}
 \beta_0 = -1 \quad , \quad \gamma = -2  \quad , \quad \rho_0 = -q_0 > 0  
 \quad , \quad  z_S >0 \quad, 
\end{equation}
and the metric (\ref{eq:MetricAffine1}) turns into
\begin{eqnarray}\label{eq:MetricAffineMaxwell}
 g_{Maxwell} & = & \left( 1- \frac{3\, z}{3\, z_S- 1} \right)^{ -2}
 \, \left( 1 - \frac{z}{z_S} \right) \,  \, dt^2 \nonumber \\
 &-&
 \, \left( 1- \frac{3\, z}{3\, z_S- 1} \right)^{2}
 \, \left( 1 - \frac{z}{z_S} \right)^{-2}  \, (dx^2 + dy^2) \\
 &-&
 \, \frac{3}{z_s \, \rho_0 \, (3 \, z_S - 1)^2} 
 \, \left( 1- \frac{3\, z}{3\, z_S- 1} \right)^{2}
 \, \left( 1 - \frac{z}{z_S} \right)^{-5}  \, dz^2 \nonumber
\end{eqnarray}
If we set  
\begin{equation}
  z = \frac{a - 1}{3} \, \left(  \frac{a \, \sigma \, w}{a \, \sigma \, w \, + \, 1} 
  \right) \qquad , \qquad a = 3\, z_S \quad , \quad 
  \sigma =  \sqrt{\frac{\rho_0}{a}} \quad ,
\end{equation}
the metric (\ref{eq:MetricAffineMaxwell}) is represented in coordinates for which the electric 
field is uniform: 
\begin{equation}\label{eq:MetricAffine1Maxwell}
\begin{array}{c}
 g_{Maxwell}  = (\,1 + a \, \sigma \, w \,)^{} \, (\,1 + \sigma \, w \,)^{} \, dt^2 
 - \, (\,1 + \sigma \, w \,)^{-2}  \, (dx^2 + dy^2) \\[2mm]
 - (\,1 + a \, \sigma \, w \,)^{-1} \, (\,1 + \sigma \, w \,)^{-5} \, dw^2 \quad .
\end{array}
\end{equation}
(See formula (59) in \cite{AmGr} and references therein. )%

 We can also consider a {\bf perfect fluid} ($p=q$) with cosmological constant and the prescribed 
equation of state 
\begin{equation}\label{eq:RelacaoLinearPerfFluid}
 \rho + \Lambda = \beta_0 \, (p - \Lambda) \qquad 
\end{equation}
This corresponds to setting $\gamma = \beta_0 + 7$ in the metric (\ref{eq:MetricAffine1}). 
Such solutions have been studied at least from the 1950's (\cite{Taub}) and are still
matter of interest (\cite{Sar}). For more references, the reader could consult 
\cite{TWS}, \cite{BK}, \cite{Col} and \cite{SKMHH}. 


%
\item[{\bf (ii)}] {\bf \, Linear System II ($\gamma \neq 0$ , $\beta_0 = 1$) :}
 Here the function $q(z)$ is given as 
\begin{equation}\label{eq:qAffine2}
 q = \Lambda + \, (q_0 - \Lambda) 
 \, \left( 1 - \frac{z}{z_S} \right)^{1- \frac{6}{\gamma}} 
  \, e^{- \frac{6}{\gamma} \, z} 
\end{equation}
and the metric as
\begin{eqnarray}\label{eq:MetricAffine2}
 g &=& \, \left( 1 - \frac{z}{z_S} \right)^{-\frac{2}{\gamma}} \,
 e^{\frac{6}{\gamma}\, z}  \, dt^2 
 -
 \, \left( 1 - \frac{z}{z_S} \right)^{\frac{4}{\gamma}} \, (dx^2 + dy^2) \\
 &-&
 \, \frac{12}{z_S \, (\Lambda - q_0)\, \gamma^2} 
 \, \left( 1 - \frac{z}{z_S} \right)^{- 2 + \frac{6}{\gamma}} 
  \, e^{\frac{6}{\gamma}\, z} \, dz^2 \quad . \nonumber
 \end{eqnarray} 
The constant $z_S$ is chosen such that $z_S \,  (\Lambda - q_0) > 0$ .
As a special example, if $\gamma = 8$ and $\Lambda = 0$ we have 
the {\bf perfect fluid} solution ($p=q$) with the equation of state $\rho = p$ :
\begin{eqnarray}\label{eq:MetricAffine2PrefFluid}
 g_{PF} &=& \, \left( 1 - \frac{z}{z_S} \right)^{-\frac{1}{4}} \,
 e^{\frac{3}{4}\, z}  \, dt^2 
 -
 \, \left( 1 - \frac{z}{z_S} \right)^{\frac{1}{2}} \, (dx^2 + dy^2) \\
 &-&
 \, \frac{3}{ (- \, z_S \, \rho_0)\, 16} 
 \, \left( 1 - \frac{z}{z_S} \right)^{- \frac{5}{4}} 
  \, e^{\frac{3}{4}\, z} \, dz^2 \quad ,\nonumber
\end{eqnarray} 
with $z_S \, \rho_0 < 0$. Rescaling $t,x,y$ by 
suitable constant parameters and defining
\begin{equation} \label{eq:coordinatePF}
 w = \alpha \,  \left( 1 - \frac{z}{z_S} \right)^{\frac{1}{2}} \, ,
 \qquad  \alpha =  
 \left(- \,  \frac{3\,z_S \, e^{\frac{3\,z_S}{4}} }{4 \, \rho_0} \right)^{\frac{2}{3}}\, ,
 \quad   \kappa = \frac{\sqrt{-3 \, \rho_0 \, z_S }}{2 \, \alpha}  
\end{equation}
the metric (\ref{eq:MetricAffine2PrefFluid}) turns into the Tabensky-Taub solution 
(\cite{TaTau})
\begin{eqnarray}\label{eq:MetricAffine2PrefFluid2}
 g_{PF} &=& \frac{e^{\frac{(\kappa \, w)^2}{\rho_0}}}{\sqrt{w}}  \,( dt^2 - dw^2 ) 
 -
 \, w \, (dx^2 + dy^2) \quad .
\end{eqnarray}

%
\item[{\bf (iii)}] {\bf \, Linear System III ($\gamma = 0$ , $\beta_0 \neq 1$) : }
In this case
\begin{equation}\label{eq:qAffine2}
 q = \Lambda + \, (q_0 - \Lambda) 
 \, \left( 1 - \frac{z}{z_S} \right)^{-\frac{\beta_0 + 1}{\beta_0 -1}} 
  \, e^{\, \frac{2}{z_S \, (\beta_0 - 1)} \, \frac{z}{z-z_S}} \quad .
\end{equation}
and
\begin{equation}\label{eq:MetricAffine2}
 \begin{array}{c}
 g = \,  \left( 1 - \frac{z}{z_S} \right)^{\,\frac{2}{\beta_0 -1}} 
  \, e^{\, \frac{2}{3 z_S \, (\beta_0 - 1)} \, \frac{z}{z-z_S}} \, dt^2 
 -
 \, e^{\, - \, \frac{ 4}{3 z_S \, (\beta_0 - 1)} \, \frac{z}{z-z_S}} \, (dx^2 + dy^2) \\[6mm]
 -
 \, \frac{4}{3 z_S^3 \, (\Lambda - q_0)\, (\beta_0 - 1)^2} \, 
 \left( 1 - \frac{z}{z_S} \right)^{\frac{\beta_0 + 1}{\beta_0 -1} - 3} 
  \, e^{- \, \frac{2}{ z_S \, (\beta_0 - 1)} \, \frac{z}{z-z_S}}  \, dz^2
 \end{array} 
\end{equation}
with $z_S$ is chosen such that $z_S \,  (\Lambda - q_0) > 0$ .

Here, if we set $\Lambda = 0$ and $\beta_0 = -7$ we find the remaining and unphysical
perfect fluid solution ($p=q$) with $\rho = - 7 \, p$.

\end{enumerate}


\section{The Special Solutions}

 If the energy-momentum tensor is not generic then there must exist 
an open interval around $z=0$ where %
\footnote{$\rho = -p = q - 2 \Lambda$ is equivalent to $\rho - q + 2 \Lambda = 0 $ and 
           $\rho + 4 p + 3 q -6 \Lambda = 0$.}
\begin{equation}
 q = \Lambda \qquad \text{or} \qquad
 \rho = -p = q - 2 \Lambda \quad .
\end{equation}
In this section we analyze these two possibilities. 

\begin{thm}[Solution with $q=\Lambda$]~ \label{thm:GeneralSolution2} \\
Assuming there is an open interval $I_0$ containing $z=0$ where 
\begin{equation}
 q = \Lambda \quad , 
\end{equation}
and that the metric $g$ satisfies Einstein's equations with a cosmological 
constant $\Lambda$, then one of the following 
relations will be satisfied for a suitable choice of the coordinate function $z$ :
\begin{enumerate}
\item[(a)] For every $z \in I_0$,    $\rho(z) = - \Lambda$ and  
\begin{equation}\label{eq:GeneralSolutionMetric2RhoIgualLambda}
 g = e^{2 \int_0^z dz' \, \frac{z'-z_0}{p(z')-\Lambda - (z'-z_0)^2}} \, dt^2
 - \, \left(dx^2 + dy^2 \right) - \, \frac{dz^2}{(\, p(z)-\Lambda - (z-z_0)^2 \, )^2}  \quad ,
\end{equation}
where $z_0$ is an arbitrary constant chosen such that $p_0 -\Lambda - (z_0)^2 \neq 0$.
%
%

\item[(b)] For every $z \in I_0$,  $\rho(z) =-4\, p(z) + 3 \Lambda$  and 
\begin{equation}\label{eq:GeneralSolutionMetric2Segunda}
 g = e^{-{2 \over 3} \int_0^z dz' \, \frac{z'-z_0}{\Psi(z')}} \, dt^2
 - e^{{4 \over 3} \int_0^z dz' \, \frac{z'-z_0}{\Psi(z')}} \, \left(dx^2 + dy^2 \right) 
 -  \,\left(\, \frac{dz}{\Psi(z)}\, \right)^2 \quad ,
\end{equation}
where $z_0$ is an arbitrary constant chosen such that $\Psi(0) \neq 0$ and
\begin{equation}\label{eq:Psi}
\Psi(z)  =  3 \, ( \, p(z)-\Lambda \, )\,  - (z-z_0)^2 \, \quad .
\end{equation}
\end{enumerate}
\end{thm}

\proof
If $q=\Lambda$, then from equation (\ref{Eq:EinsteinDolgov3}) we conclude that $\varphi = 0$ 
or $\psi =0$. 
\begin{enumerate}
\item[(a)] If $\varphi =0$, then $\rho = - \Lambda$, from (\ref{Eq:EinsteinDolgov1}) . 
Therefore   
\begin{equation}
 g = e^{2 \int\, du \, \psi(u)} \, dt^2
 - \, \left(dx^2 + dy^2 \right) - \, du^2 \quad , 
\end{equation}
with $\psi$ satisfying equation (\ref{Eq:EinsteinDolgov2}):
\begin{equation}
d\psi  =\left( \, p - \Lambda -  \psi^2 \, \right) \, du \quad .
\end{equation}
Defining the "new" coordinate function as $z=\psi + z_0$, with $z_0$
an arbitrary constant, we obtain (\ref{eq:GeneralSolutionMetric2RhoIgualLambda}).
\item[(b)]  If $\psi =0$, then $\rho = -4p + 3\Lambda$, from
(\ref{Eq:EinsteinDolgov2}). Therefore 
\begin{equation}
 g = e^{-{2 \over 3} \int du \, \varphi(u)} \, dt^2
 - e^{{4 \over 3} \int du \, \varphi(u)} \, \left(dx^2 + dy^2 \right) - \, du^2 \quad , 
\end{equation}
with $\varphi(u)$ satisfying the equation
\begin{equation}
d\varphi  = \left( \, 3(p - \Lambda) -  \varphi^2 \, \right) \, du \quad .
\end{equation}
Defining the "new" coordinate function as $z=\varphi + z_0$, with $z_0$
an arbitrary constant, we obtain (\ref{eq:GeneralSolutionMetric2Segunda}).
\end{enumerate}
\qed
%
%

\begin{thm}[Solution with $\rho  = -p = q - 2 \Lambda$]~ \label{thm:GeneralSolution3} \\
Assuming there is an open interval $I_0$ containing $z=0$ where 
\begin{equation}
\rho  = -p = q - 2 \Lambda \quad , 
\end{equation}
with $( t,x,y,z )$ a coordinate system adapted to the symmetry and for which $g_{33} =- 1$,
and that the metric $g$ satisfies Einstein's equations with a cosmological constant $\Lambda$, 
then there are constants $\alpha$ and $\beta$ such that  
for every $z \in I_0$ 
\begin{equation}\label{eq:generalsolutionq3}
 q =  \Lambda + \frac{4 \, \alpha \, \beta }{3 \,(1+(\, \alpha + \beta \,) \, z \,)^2}
\end{equation}
and,  if $\alpha + \beta \neq 0$,\vspace{1mm}
\begin{equation}\label{eq:GeneralSolutionMetric3}
 g = \left(1 + (\, \alpha + \beta \, ) \, z\, \right)
 ^{\, \frac{2}{3}\left( \frac{3 \alpha - \beta}{\alpha + \beta} \right)} \, dt^2
 - \, \left(1 + (\, \alpha + \beta \, ) \, z\, \right)
 ^{\, \frac{4}{3}\left( \frac{\beta}{\alpha + \beta} \right)} \, \left(dx^2 + dy^2 \right) 
 - \, dz^2 \quad , 
\end{equation}
\vspace{1mm} 
or, if $\alpha + \beta = 0$, 
\vspace{1mm}
\begin{equation}\label{eq:GeneralSolutionMetric32}
 g = e^{\, \frac{8}{3} \, \alpha \, z} \, dt^2
 - \, e^{\, - \, \frac{4}{3} \, \alpha \, z} \, \left(dx^2 + dy^2 \right) 
 - \, dz^2 \quad .
\end{equation}
\vspace{1mm}
In the special case $q = \Lambda$ we obtain, for $\beta = 0$, the Minkowski metric  described 
by an observer with a uniform acceleration $\alpha$ (\cite{GrHer}\,) 
or, for $\alpha = 0$, the Taub-Levi-Civita vacuum solution (\cite{AmGr}\,) . 
\end{thm}
%
%
\proof
Applying the hypothesis of the theorem in equations 
(\ref{Eq:EinsteinDolgov1})-(\ref{Eq:EinsteinDolgov3}), we obtain the following system 
of ODE's:
\begin{equation}
 \varphi' + \varphi^2 + \psi \, \varphi = 0  \qquad 
 \psi' + \psi^2  + \psi \, \varphi = 0  \quad . 
\end{equation}
Its general solution is, after defining the "new" coordinate function $z=u$,
\begin{equation}
 \psi = \frac{\alpha }{1 + (\alpha + \beta) \, z}  \qquad 
 \varphi = \frac{\beta}{1 + (\alpha + \beta) \, z}  \quad . 
\end{equation}
Applying them in the metric (\ref{eq:SolutionMetricDogov}) we obtain (\ref{eq:GeneralSolutionMetric3})
and (\ref{eq:GeneralSolutionMetric32}).
Using (\ref{Eq:EinsteinDolgov3}) we get (\ref{eq:generalsolutionq3}).
\qed
%
%


\section{Concluding remarks}

 Three possible "types" of solutions to static plane symmetric 
Einstein's equations with cosmological constant have been given. 
It remains to show that they cover any possible solution, that is, 
locally there are coordinates for which the metric takes the form 
as in one of the three theorems presented so far.  

 Any solution with a non-generic energy-momentum tensor has a local
representation like at least one among those given in theorems \ref{thm:GeneralSolution2} and 
\ref{thm:GeneralSolution3}, as it is clear from their proofs.
Further explanation is necessary for a generic energy-momentum tensor . In order to do so,
define for the metric (\ref{eq:SolutionMetricDogov}) the "new" coordinate $z$ as
\begin{equation}\label{eq:FinalRemarkszu}
 z= z_S + \frac{\psi(u)}{\varphi(u)} \quad.
\end{equation}
The inversion function theorem in its simplest form tell us that this is a good 
coordinate definition as far as $\frac{du}{dz}(0) \neq 0$. 
Assuming that Einstein's equations 
(\ref{Eq:EinsteinDolgov1})-(\ref{Eq:EinsteinDolgov3}) hold
and expressing the results in terms of $z$ and $\Phi(z)$, we find
\begin{equation}\label{eq:FinalRemarkszu2}
 \varphi(u) \, \frac{du}{dz} = 3 \, \Phi(z) 
 \qquad \text{and} \qquad
 \psi(u) \, \frac{du}{dz} = 3 \,(\, z - z_S \, )\,  \Phi(z) \quad .
\end{equation}
Hence the coordinate transformation is well defined as far as the energy-momentum
tensor is generic. Substituting these identities in the metric (\ref{eq:SolutionMetricDogov})
we get exactly the formula given in theorem \ref{thm:GeneralSolution}. 

We conclude that any solution 
of static plane symmetric Einstein's equations with cosmological constant has a 
local  behavior as stated in one of the three theorems we have considered in this paper.



\begin{thebibliography}{99}

\bibitem{AmGr} \textsc{P. A. Amundsen, \O{}. Gr\o{}n:}
 \emph{General static plane-symmetric solutions of the Einstein-Maxwell equations},
 Phys. Rev. D , Vol. 27, {\bf 8}, 1983.

\bibitem{BK} \textsc{K. A. Bronnikov, M. A. Kovalchuk:}
 \emph{Properties of static fluid cylinders and plane layers in general relativity},
 Gen. Rel. Grav. {\bf 11}, 1979.

\bibitem{Col} \textsc{C. B. Collins :}
 \emph{ Static relativistic perfect fluids with spherical, plane, or hyperbolic symmetry},
 J. Math. Phys. {\bf 26} (9), September 1985.

\bibitem{DolgKhri} \textsc{A. D. Dolgov, I. B. Khriplovich:}
 \emph{Does a Static Solution Exist for a Gravitating Planar Wall?},
 Gen. Rel. and Grav. , Vol. 21, {\bf 1}, 1989.

\bibitem{GrHer} \textsc{\O{}. Gr\o{}n, S. Hervik:}
 \emph{Einstein's General Theory of Relativity},
 Springer, 2007.
 
\bibitem{Hors} \textsc{J. Horsk\'y :}
 \emph{The gravitational field of a homogeneous plate with a non-zero cosmological 
 constant}, Czech. J. Phys. B , {\bf 25}, 1975.

\bibitem{McV} \textsc{G. C. McVittie :}
 \emph{On Einstein's Unified Field Theory}, Proc. R. Soc. London A, {\bf 124}, 1929.

 
\bibitem{NoHors} \textsc{J. Novotn\'y, J. Horsk\'y :}
 \emph{On the plane gravitational condensor with the positive gravitational constant}, 
 Czech. J. Phys. B , {\bf 24}, 1974.
 
 

\bibitem{Sar} \textsc{R. E. G. Sarav\'i  ́:}
 \emph{ Static plane symmetric relativistic fluids and empty repelling singular 
 boundaries}, Class. Quantum Grav. , {\bf 25},  2008

\bibitem{SKMHH} \textsc{H. Stephani, D. Kramer, M. Maccallum, C. Hoenselaers, E. Herlt:}
 \emph{Exact Solutions to Einstein's Field Equations}, Second Edition,
 Cambridge , 2003.

\bibitem{TaTau} \textsc{R. Tabensky, A. H. Taub :}
 \emph{Plane Symetric Self-gravitating Fluids with Preassure Equal to Energy Density}, 
 Commun. Math. Phys. Rev. , {\bf 29}, 1973.

 \bibitem{Taub} \textsc{A. H. Taub :}
 \emph{Isentropic hydrodynamics in plane symmetric space-times}, 
 Phys. Rev. , vol. 103, {\bf 2}, 1956.
 
\bibitem{TWS} \textsc{A. F. F. Teixeira, I. Wolk, M. M. Som :}
 \emph{Exact relativistic solution of disordered radiation with planar symmetry}, 
 J. Phys. A , {\bf 10}, 1977.
 
\end{thebibliography}
\end{document}